\documentclass{article}

\usepackage{PRIMEarxiv}

\usepackage[utf8]{inputenc} % allow utf-8 input
\usepackage[T1]{fontenc}    % use 8-bit T1 fonts
\usepackage{hyperref}       % hyperlinks
\usepackage{url}            % simple URL typesetting
\usepackage{booktabs}       % professional-quality tables
\usepackage{amsfonts}       % blackboard math symbols
\usepackage{nicefrac}       % compact symbols for 1/2, etc.
\usepackage{microtype}      % microtypography
\usepackage{lipsum}
\usepackage{fancyhdr}       % header
\usepackage{graphicx}       % graphics

\usepackage{booktabs} % MAG
\usepackage{multirow} % MAG
\newcommand{\otoprule}{\midrule[\heavyrulewidth]} % MAG

\graphicspath{{media/}}     % organize your images and other figures under media/ folder

\title{Deep Neural Network for Automatic Assessment of Dysphonia
}

\author{
  Mario Alejandro García\\
  Universidad Tecnológica Nacional \\
  Facultad Regional Córdoba\\ 
  Argentina\\
  \texttt{mgarcia@frc.utn.edu.ar} \\
   \And
  Ana Lorena Rosset \\
  Escuela de Fonoaudiología\\
  Universidad Nacional de Córdoba \\
  Argentina\\
  \texttt{lorena.rosset@unc.edu.ar} \\
}

\begin{document}
\maketitle

\begin{abstract}
The purpose of this work is to contribute to the understanding and improvement of deep neural networks in the field of vocal quality. A neural network that predicts the perceptual assessment of overall severity of dysphonia in GRBAS scale is obtained. The design focuses on amplitude perturbations, frequency perturbations, and noise. Results are compared with performance of human raters on the same data. Both the precision and the mean absolute error of the neural network are close to human intra-rater performance, exceeding inter-rater performance. 
\end{abstract}

% keywords can be removed
% \keywords{First keyword \and Second keyword \and More}


\section{Introduction}

Auditory-perceptual judgment is a main part of routine clinical assessment of patients with voice disorders to document the voice quality \cite{barsties2015assessment}. The assessment of voice quality is important to detect both vocal pathologies and other diseases that, although they do not originate in the vocal cords, show signs of impaired vocal quality, for example Parkinson's disease \cite{hazan2012early, little2008suitability}. This assessment is also important in monitoring the treatment.

\subsection{GRBAS}

In order to standardize the evaluation and interrelate the auditory and physiological aspects of vocal production, methods and scales of audioperceptive evaluation have been proposed, such as GRBAS and CAPE-V \cite{barsties2015assessment}. 

The GRBAS scale is an audioperceptive voice assessment method. It is based on studies that began in 1966 by the Japan Society of Logopedics and Phoniatrics \cite{isshiki1966approach} and was later popularizated by Minoru Hirano in 1981 \cite{hirano1986clinical}. The scale was globally adopted and it was also validated in many countries \cite{yun2005correlation, hui2007validation, karnell2007reliability, jesus2009voice}. It is currently used by both clinicians and researchers. 

GRBAS comprises five dimensions (which form the acronym GRBAS) for the assessment of the glottal source: Grade or overall severity of dysphonia, Roughness, Breathiness, Asteny and Strain.

Each dimension is rated on an integer four point scale, from ``0'' (no dysphonia) to ``3'' (severe dysphonia).

There are two major weaknesses in the auditory-perceptual methods, the subjectivity of voice assessment and the need for experienced listeners \cite{kreiman2010perceptual, nunez2012espectrograma}.

\subsection{Variability in quality assessment}

Even using scales such as GRBAS, the assessment of voice quality has great variability between health professionals (inter-rater variability). The same can be found between different assessment instances carried out by the same evaluator (intra-evaluator variability) \cite{gordillo2018hitos}. 

As Kreiman and Gerratt explain, ``Voice quality is an interaction between an acoustic voice stimulus and a listener; the acoustic signal itself does not possess vocal quality, it evokes it in the listener'' \cite{kreiman1998validity}. Audioperceptive assessment is affected by multiple factors. As a consequence, the level of intra-rater agreement and, even more so, the level of inter-rater agreement, can be very poor \cite{kreiman2010perceptual}. Some of the factors that affect inter-rater variability are the listener's bias regarding knowledge of the medical diagnosis, the experience of the rater, professional training (for example, in otorhinolaryngology, speech therapy, voice teachers, phoneticians and students of undergraduate or postgraduate), musical training and training in auditory perceptual judgment (for example, of native listeners) \cite{gordillo2018hitos}.

The factors mentioned in the previous paragraph cannot explain intra-rater variability, but even so, under the same conditions, different assessment instances can result in different values. Then, it can be stated that there is no ``correct answer'' for the assessment of a voice, but rather multiple assessments with different results and a certain degree of agreement. 

\subsection{Automatic classification of vocal quality}

The automatic classification of vocal quality in the audioperceptive assessment scales could reduce the subjectivity of the current analysis, standardizing the assessment among different raters and allowing repeatable results for the same rater. This is an active research topic. In the classical approach, a set of acoustic measurements are taken as features in order to fit some machine learning model.

Deep learning is the state of the art in most pattern recognition tasks. Audio pattern recognition is no exception, although significant advances have been made in speech recognition. Vocal quality and the detection/classification of pathologies have been little addressed with this technology. Some of the published works use neural models that were originally developed for image recognition \cite{huang2017densely, muhammad2018edge, alhussein2018voice}. In other works only dense layers are used \cite{xie2016deep, fang2019detection}. Few \emph{ad hoc} networks have been presented. In the category of \emph{ad hoc} models are those developed by Arias-Lodoño \emph{et al.} \cite{arias2019multimodal} and Harar \emph{et al.} \cite{harar2017voice}.

This work aims to contribute to the understanding and improvement of deep neural networks applied to the classification of vocal quality. To achieve this goal, a neural network that classifies the grade of dysphonia G was developed. The network, whose input is the power cepstrum, was designed to extract information similar to some acoustic measures related to voice quality.

\section{Materials and methods}

\subsection{Evaluation} \label{evaluacion}

\noindent The lack of an unquestioned reference and the impossibility of mapping a voice to a quality level without error through a function, however complex \cite{kreiman1996perceptual}, complicates both the training and the evaluation of the machine learning model. 

Given the assumption that the ``correct answer'' exists, the variations must be considered noise, where noise is a random value whose distribution depends on both the acoustic signal and the rater. This random component implies that it is impossible to achieve an automatic classification that exactly agrees with a particular human classification on a given data set. There is an irreducible error in the classification. Consequently, in order to measure the performance of an automatic voice quality classification system, its accuracy must be analyzed according to the context of the database (voices + assessments). If the deviation between the model's prediction and the rater's assessment is close to the intra-rater variability, it can be said that the model has a near-human assessment capability. 

In this work, the expected output on the training data is the assessment done by an individual rater. Then, the variation between the neural network results and the expected outputs are compared with the intra-rater and inter-rater variations of human experts on the same data. The mean intra-rater agreement in G is equal to 73\% for the data used in this work and 53\% for the inter-rater case. 

Table \ref{tab:intro_cm1} shows, as an example with the dataset used in this work, the confusion matrix (or contingency table) between the first assessment of raters 1 and 2. Table \ref{tab:intro_cm2} shows the confusion matrix for the first and second assessment of rater 1. 

\begin{table}[ht]
	\caption[Matriz de confusión interevaluador para datos PVQD]{\label{tab:intro_cm1}Confusion matrix between the first assessment of raters 1 and 2.}
	\centering
        \begin{tabular}{@{}cc|cccc@{}}
            \multicolumn{1}{c}{} &\multicolumn{1}{c}{} &\multicolumn{4}{c}{Rater 2} \\ 
            \multicolumn{1}{c}{} & 
            \multicolumn{1}{c|}{} & 
            \multicolumn{1}{c}{0} & 
            \multicolumn{1}{c}{1} & 
            \multicolumn{1}{c}{2} & 
            \multicolumn{1}{c}{3} \\ 
            % \cline{2-6}
            \cmidrule{2-6}
            \multirow[c]{4}{*}{\rotatebox[origin=tr]{90}{Rater 1}}
            & 0 & 48 & 7 & 1 & 0   \\
            & 1 & 31 & 30 & 17 & 0 \\ 
            & 2 & 5  & 16 & 10 & 3 \\             
            & 3 & 0  & 5 & 19 & 14 \\             
%            \cline{2-4}
        \end{tabular}
\end{table}

\begin{table}[ht]
	\caption[Matriz de confusión intraevaluador para datos PVQD]{\label{tab:intro_cm2}Confusion matrix for the first and second assessment (ASMT) of rater 1.}
	\centering
        \begin{tabular}{@{}cc|cccc@{}}
            \multicolumn{1}{c}{} &\multicolumn{1}{c}{} &\multicolumn{4}{c}{ASMT 2} \\ 
            \multicolumn{1}{c}{} & 
            \multicolumn{1}{c|}{} & 
            \multicolumn{1}{c}{0} & 
            \multicolumn{1}{c}{1} & 
            \multicolumn{1}{c}{2} & 
            \multicolumn{1}{c}{3} \\ 
            % \cline{2-6}
            \cmidrule{2-6}
            \multirow[c]{4}{*}{\rotatebox[origin=tr]{90}{ASMT 1}}
            & 0 & 39 & 17 & 0 & 0   \\
            & 1 & 19 & 49 & 8 & 2 \\ 
            & 2 & 1  & 9 & 20 & 4 \\             
            & 3 & 0  & 0 & 4 & 34 \\             
%            \cline{2-4}
        \end{tabular}
\end{table}

The model to be presented classifies the grade of dysphonia G. Performance is expected to be close to intra-rater human performance on the accuracy and mean absolute error (MAE) metrics. 

G is an ordinal categorical variable. The values of G represent a hierarchy and therefore not all disagreements are of equal importance. For instance, a disagreement between the categories ``0'' and ``3'' is clearly greater than a disagreement between ``0'' and ``1''. The MAE metric takes account of this difference. It is calculated on the numeric values representing the class names, ie 0 for class ``0''. 

In the data used, the inter-rater MAE is 0.52 and 0.28 for the intra-rater case. The database contains six assessments made by three different professionals for each audio plus an extra assessment per audio that was added during this work. 

\subsection{Data}

\subsubsection{Database}

The Perceptual Voice Qualities Database\footnote{\url{https://voicefoundation.org/health-science/videos-education/pvqd/}} (PVQD) is a public database provided by The Voice Foundation. Contains voice samples rated by experienced professionals. It contains 296 audio files consisting of the sustained /a/ and /i/ vowels, as well as continuous speech sentences defined by the Consensus Auditory-Perceptual Evaluation of Voice (CAPE-V). Audio is encoded in 16 bits and a sample rate of \mbox{44.1 KHz}. Audio files have been edited to remove clinical instructions, but some sounds persist and contaminate the samples, therefore it is necessary to review, edit and make decisions about each file in order to use this data in a machine learning process. In addition to audio data, information is provided on gender, age, diagnosis and assessments in GRBAS and CAPE-V scales. Four voice healthcare professionals rated the recordings, although the last one (rater 4) assessed only 16\% of the cases. Each professional rated each audio twice. 
More details on PVQD can be found in \cite{walden2020perceptual}. 

Table \ref{tab:db_pvqd_G} shows the number of cases per value of G for each rater and each assessment instance. 

\begin{table}[ht]
	\caption[PVQD, valoraciones de G]{Grade of dysphonia G by rater and assessment instance in PVQD.}\label{tab:db_pvqd_G}
	\centering
	\begin{tabular}{ccrrrrr}
		\toprule
		\multicolumn{2}{c}{} & \multicolumn{4}{c}{G}\\
% 		\cline{3-6}
        \cmidrule{3-6}
		Rater & Instance & \multicolumn{1}{c}{0} & \multicolumn{1}{c}{1} & \multicolumn{1}{c}{2} & \multicolumn{1}{c}{3} & Total \\
	    \otoprule
         1 & 1 & 79 & 119 & 48 & 50 & 296\\
         1 & 2 & 86 & 111 & 48 & 51 & 296\\
         2 & 1 & 113 & 98 & 61 & 24 & 296\\
         2 & 2 & 114 & 96 & 62 & 24 & 296\\
         3 & 1 & 137 & 71 & 52 & 36 & 296\\
         3 & 2 & 120 & 85 & 55 & 36 & 296\\
         4 & 1 & 14 & 27 & 8 & - & 49\\
         4 & 2 & 21 & 23 & 4 & 1 & 49\\		
         \bottomrule
	\end{tabular}
\end{table}

\subsubsection{Data preprocessing}

First, the fourth rater assessments were removed.

The author of PVQD reports that the raters have sufficient experience in the voice area, but does not provide information on which of them is the most experienced, or how much experience they have. This data would be useful, both to trust the PVQD assessments, and to decide which set of assessments to use during the training and evaluation of the neural network. Because the PVQD audios must be checked and it is desirable to have more information regarding the rater, the complete database was analyzed and evaluated by a local professional with more than 20 years of experience in clinical voice treatments. To this end, a cooperation agreement was signed with the \emph{Escuela de Fonoaudiología}\footnote{\url{https://fono.fcm.unc.edu.ar/}} of the \emph{Facultad de Ciencias Médicas} of the \emph{Universidad Nacional de Córdoba}. The steps carried out to check and assess the PVQD audios are detailed below.

\begin{enumerate}

\item Sustained /a/ vowel segmentation. The audio files consist of sustained /a/ and /i/ vowels, continuous audio phrases and, in some cases,  glissandos. These elements are not always in the same order, do not have the same length, nor do they all exist in all files. In this step, all the audio files were manually edited with Audacity\footnote{\url{https://www.audacityteam.org}} software, keeping only the sustained vowel /a/. The original file names and encoding were kept. Audio files shorter than one second were removed. 

\item File renaming and spreadsheet creation. File names have some relation to the vocal quality. There are groups of files that begin with the same letters and the vocal quality level is different among these groups. In order to avoid bias in the new assessment, random names were generated. A form (spreadsheet) was provided to the new rater. In the spreadsheet audios were ordered alphabetically by the new names. 

\item Analysis and assessment. In a noise-free environment, with Sennheiser HD 202 headphones, a PC with external sound card (Audiobox PreSonus), Audacity software and taking care of the recommended rest times, the local rater analyzed and assessed the audio on the GRBAS scale. Whenever the professional considered that an audio was not suitable for evaluation, she wrote it down in the form and wrote a comment. Whenever necessary, a comment or recommendation was written.

\item Selection and correction of audio files. Files flagged by the local professional were removed and comments were analyzed. For the files with comments in the spreadsheet, the decision was made to delete, edit or keep them in their original state as appropriate. 

\item Creation of transactional database. The spreadsheet data was loaded into a SQLite\footnote{\url{https://www.sqlite.org}} database. GRBAS assessments of PVQD were also inserted.

\end{enumerate}

In the first step, 6 audio files were removed because they did not meet the minimum length condition. 

Table \ref{tab:exp_GRBAS_ALR} summarizes the assessments obtained in step 3. NC column contains the number of files that could not be assessed in each GRBAS category. The reasons are ambient noise, unstable voices (segments with different assessment value), and singing voices with vibrato (intentional periodic variation of the fundamental frequency). Non-assessed files were flagged for deletion.

\begin{table}[ht]
	\caption[Valoración local de la PVQD]{Assessment of local rater.}
	\label{tab:exp_GRBAS_ALR}
	\centering
	\begin{tabular}{crrrrrr}
		\toprule
		\multicolumn{1}{c}{} & \multicolumn{4}{c}{Value}\\
% 		\cline{2-5}
        \cmidrule{2-5}
		Dimension & 
		\multicolumn{1}{c}{0} & \multicolumn{1}{c}{1} & \multicolumn{1}{c}{2} &
		\multicolumn{1}{c}{3} &
		\multicolumn{1}{c}{NC} & Total \\
		
	    \otoprule
        G & 61 & 126 & 65 & 30 & 8 & 290 \\
        R & 129 & 79 & 62 & 11 & 9 & 290 \\
        B & 59 & 139 & 65 & 18 & 9 & 290 \\
        A & 212 & 54 & 14 & - & 10 & 290 \\
        S & 238 & 24 & 13 & 5 & 10 & 290 \\
  
        \bottomrule
	\end{tabular}
\end{table}

In step 4, 17 files were removed, 2 due to vibrato, 5 due to noise, 2 due to recording errors and 8 due to non-uniformity. The deleted files had G equal to ``0'', ``1'' or ``2'', so the amount of data in the balanced data set was not changed. In addition, 8 files had their beginning or end removed to avoid noise.

\subsubsection{Data augmentation}
Deep neural networks have many internal parameters to tune, therefore they are prone to overfitting. One way to deal with overfitting is to increase the amount of training data. In this work, three transformations were performed to increase the amount of data, resampling, cropping and flipping. After the three transformations, the number of files was multiplied by 18, except when the increase in frequency could not be performed (in this case, each file was multiplied by 12). 

Before audio transformation, the recordings were downsampled to 25000 Sa/S. Downsampling was done with the \emph{resample} function of Librosa\footnote{\url{https://librosa.org}} library. 

\paragraph{Resampling}

\begin{figure*}[!t]
	\centering
	\includegraphics[width=1\textwidth]{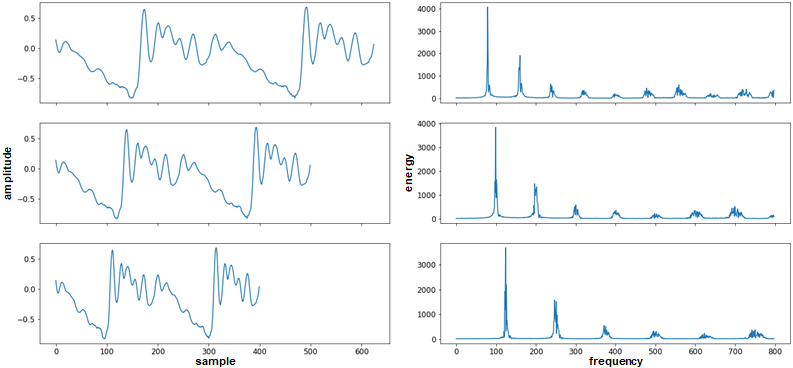}
	\caption[Aumentación de datos, desplazamiento en frecuencia]{Resampling. The original signal is shown in the middle row, in the upper row the signal decreased by 20\% in the sampling rate and in the lower row the signal with a 20\% increase. The effect in the time domain is shown in the left column and the frequency spectrum in the right column.}
	\label{fig:resamples}
\end{figure*}

Two resamples were performed for each audio file, increasing the sampling rate by 20\% (20000 Sa/S) in the first case and decreasing it by the same proportion for the second (31250 Sa/S). Each resample resulted in a new audio file, multiplying the amount of data by 3. Figure \ref{fig:resamples}  shows the results in the time and frequency domains. From top to bottom, it can be seen that the signal in the time domain is shortened, i.e. increased in frequency, while in the frequency domain the change is seen in the location and spacing of the harmonics.

Resampling to 20000 Sa/S can result in audio shorter than one second. In those cases the new signal was discarded. 

\paragraph{Cropping}

\begin{figure}[!t]
	\centering
	\includegraphics[width=.45\textwidth]{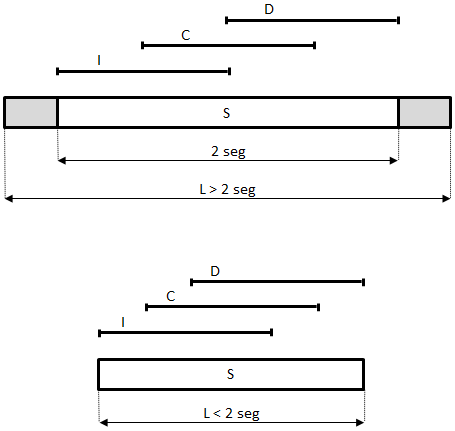}
	\caption[Aumentación de datos, segmentación por tiempo]{Cropping. Extraction of three audio segments, from a file with length $L > 2$ seconds (top) and from a file with length $L < 2$ seconds (bottom).}
	\label{fig:segmentac}
\end{figure}

Three one-second segments were extracted from each audio file. The location of the left (L), center (C) and right (R) segments depends on the length $L$ of the original audio. For $L > 2$ seconds, the middle two-second S segment was taken, otherwise the S segment was the entire original audio. Segment L is the first second of S, C is the middle second of S, and R is the last second of S. Figure \ref{fig:segmentac} shows the segmentation graphically. 

\paragraph{\emph{Flipping}}

\begin{figure*}[!t]
	\centering
	\includegraphics[width=.8\textwidth]{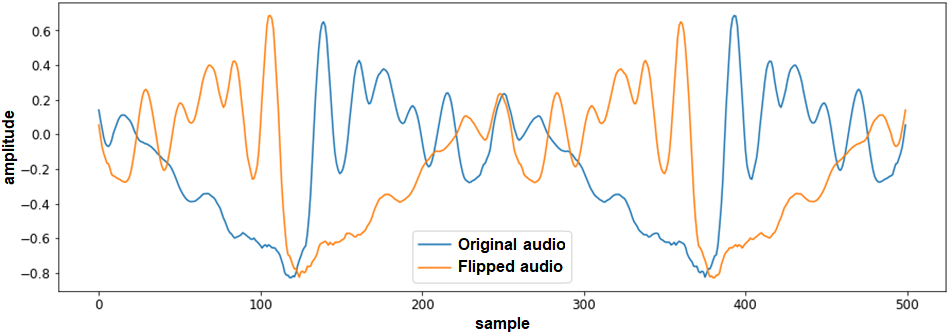}
	\caption[Aumentación de datos, \emph{flipping}]{Flipping. Original and transformed audio segments. }
	\label{fig:flipping}
\end{figure*}

A new audio was generated for each existing audio by reversing the order of the signal in time. Figure \ref{fig:flipping} shows the transformation of a small audio segment. 

\begin{figure}[!t]
	\centering
	\includegraphics[width=.45\textwidth]{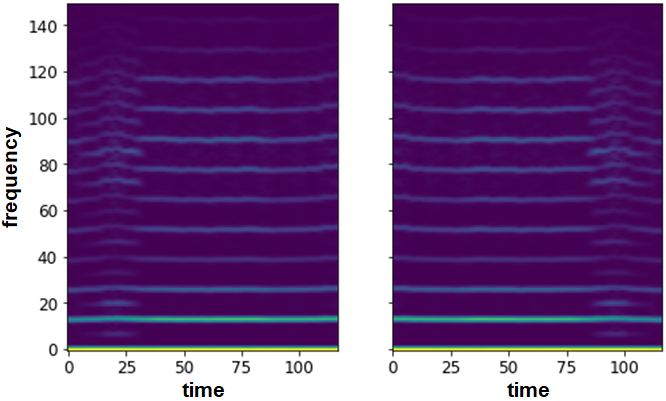}
	\caption[Aumentación de datos, \emph{flipping} (espectrogramas)]{Flipping. Spectrograms of original audio (left) and transformed audio (right).}
	\label{fig:flipping_frec}
\end{figure}

This data augmentation technique is not applicable to speech recognition because the order and shape of the phonemes are changed, but it is useful for the classification of vocal quality in sustained vowels. The frequency spectrum does not change after the transformation, but the spectrogram does. The spectrogram results in a horizontal mirroring, as can be seen in figure \ref{fig:flipping_frec}.

\subsubsection{Dataset creation}

20\% of the data was reserved for the final evaluation (test). With the remaining data, a cross-validation scheme of k iterations (k-fold) was used for k=4. In total it was necessary to create 5 same size sets. 

It is important that audios from the same person are not present in different data sets (training, validation and test), because the classifier could tend to recognize the speaker instead of G. In \cite{godino2003effects} an analysis of a similar situation (that occurs in the article \cite{ritchings2002pathological}) can be seen . To fulfill this condition, the creation of the data subsets followed this steps: 

\begin{enumerate}

\item Partitioning by origin. A table with the names of the original files and the local rater G-values was created. The table was partitioned into k parts (for k=5) called k-tables. The partitioning was done in a stratified way, that is, keeping in each k-table the same proportion of each class (G-value) as the original table. 

\item Creation of k groups of augmented data. For each k-table, a group was created containing all the audios generated in the data augmentation with the source files of the corresponding k-table.

\item Creation of k subsets. In this step, a balanced data subset was created for each group of augmented data. For each group created in the previous step, the number of occurrences of each class was calculated. The lowest amount of each group, $mc$, will be the maximum for all classes. With our data, $mc$ is always defined by the value ``3'' of G. Finally, each subset was created by randomly choosing $mc$ audios from each class in the corresponding group.   
\end{enumerate}

Five balanced data subsets were obtained, which contain a maximum of 432 audios each. The amount was reduced in some cases, although not significantly, depending on the length of the original audio files for G=``3'' (in cases where the file could not be resampled). 

\subsection{Neural network}

\begin{figure*}[!t]
	\centering
	\includegraphics[width=.66\textwidth]{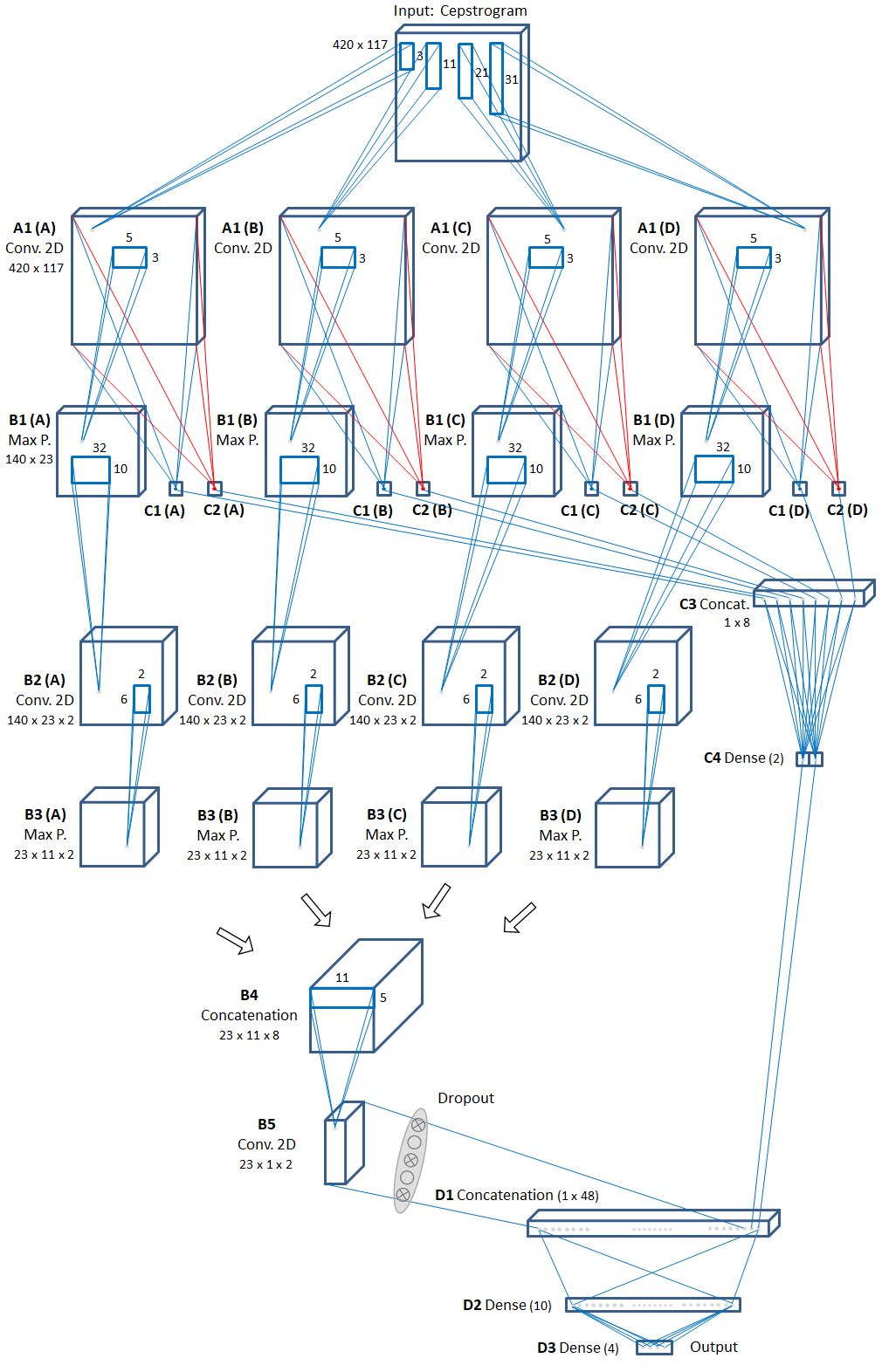}
	\caption[Modelo final]{Neural network for grade of dysphonia (G) classification.}
	\label{fig:modelo_final}
\end{figure*}

The neural network architecture was defined from an initial model based on smaller models. These small neural models, designed with knowledge of the problem domain, focus on extracting features related to amplitude perturbations, frequency perturbations and noise. Then, through an iterative process of testing, analyzing and improving the initial model, the final neural network (shown in figure \ref{fig:modelo_final}) was obtained.  

The input is the cepstrogram (a representation of the cepstrum as it varies with time) without the first 20 quefrencies to eliminate the effect of the filter (of the source-filter model of voice production). The output is G in one-hot encoding. 

Conceptually, the network can be divided into two parts, feature extraction (up to operation D1) and classification (starting at operation D1). 

The model initially performs a smoothing process of the inputs and then the extraction of features by two paths. During the optimization process, it was observed that performing different degrees or scales of smoothing, that is, more than one smoothing with windows of different width, and extracting characteristics from each scale improved the results. Smoothing is carried out in the convolution layers A1, with kernels of width 1 and height 3, 11, 21 and 31 initialized with Gaussian functions. 

One of the feature extraction paths (Path 1) was designed to obtain information related to amplitude perturbation and frequency perturbation. Path 2 was designed to obtain information related to noise. 

\paragraph{Path 1} Starts with a max pooling layer (B1) for each smoothing layer. The window of these layers has size $3 \times 5$ (height $\times$ width) and strides $3 \times 5$. Padding = ``valid'' results in an output vector of size $140 \times 23 \times 1$. The output of B1 layers is the input of convolution layers B2, with 2 kernels (each) of size $10 \times 32$, strides $1 \times 1$, padding = ``same'' and ReLU activation function. The size of the outputs of the B2 layers is $140 \times 23 \times 2$. The signal then goes through the max pooling layers B3, with windows size $6 \times 2$, strides $6 \times 2$, and \mbox{padding = ``valid''}, resulting in vectors of size $23 \times 11 \times 2$. The output of the last four max pooling layers is reshaped (B4) to $23 \times 11 \times 8$.  Finally, a new convolution (B5) is computed with 2 kernels of size $5 \times 11$, strides $1 \times 11$, padding = ``same'' and ReLU activation function. The size of the path output is $23 \times 1 \times 2$. In layers B2 and B5, L2 regularization is applied with penalty $\lambda = 0,001$. 

\paragraph{Path 2} Noise information is extracted by computing an operation similar to smoothed cepstral peak prominence (CPPS), where the linear regression is replaced by the mean. A max pooling layer (C1) and an average pooling layer (C2) are connected to each smoothing layer (A1), both with a window size $420 \times 117$ (the same size as the output of the smoothing layers), so the size of the outputs of C1 and C2 is $1 \times 1$. The four C1 and four C2 layers are then reshaped (C3) and used as input for two densely connected neurons (C4) with ReLU activation function. 

For the classification stage, both feature extraction paths join at D1. The outputs of the paths are reshaped and concatenated creating a vector of size 48. This vector is the input of a dense layer (D2) with 3 neurons and ReLU activation function, which connects to the dense layer D3, with 10 neurons and ReLU activation and. Finally, the output of D3 is the input of the four output neurons, which have sigmoidal activation. 

More information about the architecture design process can be seen at \cite{garcia2021deep}.

\section{Experimental results}
\subsection{Metrics}

The final model reaches a mean accuracy of 0.711 and MAE 0.303.

\begin{figure}[!t]
	\centering
	\includegraphics[width=.35\textwidth]{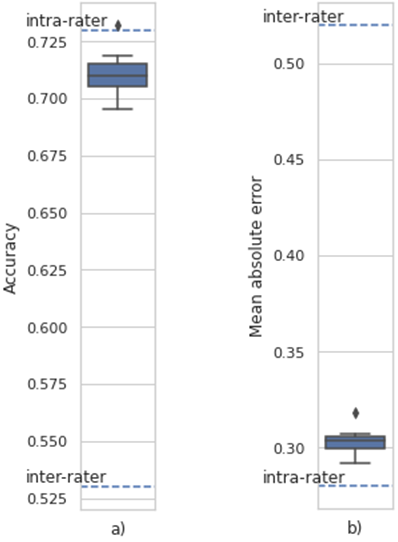}
	\caption[Resultados, métricas]{Box plot for accuracy (a) and MAE (b) in the experiments. In both cases, can be seen the comparison with the mean inter-rater and intra-rater agreement and absolute error.}
	\label{fig:res_boxplot}
\end{figure}

Figure \ref{fig:res_boxplot} shows the box plots of both metrics for 10 executions of the training process. The results are compared with the intra-rater and inter-rater agreement (joint probability of agreement) and absolute error for the same data. 

In the confusion matrix of table \ref{tab:resultados_cm}, the outputs of the classifier (first run) are compared to the local rater assessment. In this case, the accuracy was 0.713 and MAE 0.292.

\begin{table}[ht]
	\caption[Resultados, matriz de confusión evaluador/modelo]{Confusion matrix between the assessment of the local rater and the automatic assessment (first training). }\label{tab:resultados_cm}
	\centering
        \begin{tabular}{@{}cc|cccc@{}}
            \multicolumn{1}{c}{} &\multicolumn{1}{c}{} &\multicolumn{4}{c}{Neural network} \\ 
            \multicolumn{1}{c}{} & 
            \multicolumn{1}{c|}{} & 
            \multicolumn{1}{c}{0} & 
            \multicolumn{1}{c}{1} & 
            \multicolumn{1}{c}{2} & 
            \multicolumn{1}{c}{3} \\ 
            % \cline{2-6}
            \cmidrule{2-6}
            \multirow[c]{4}{*}{\rotatebox[origin=tr]{90}{Loc. rater}}
            & 0 & 67 & 41 & 0 & 0   \\
            & 1 & 14 & 70 & 24 & 0 \\ 
            & 2 & 0  & 26 & 74 & 8 \\             
            & 3 & 0  & 2 & 9 & 97 \\             
%            \cline{2-4}
        \end{tabular}
\end{table}

\subsection{Weights and hidden layer outputs}

Artificial neural networks are often considered a ``black box'' model because it is not possible to understand, at least intuitively, the relations they detect in the input data. Likewise, in some cases the analysis of the parameters obtained during the training can help in understanding. 

Unsurprisingly, the weights where it is easiest to interpret which information is useful are those of the first feature extraction filters, the weights of the B2 layers in this case.

\begin{figure*}[!t]
	\centering
	\includegraphics[width=.9\textwidth]{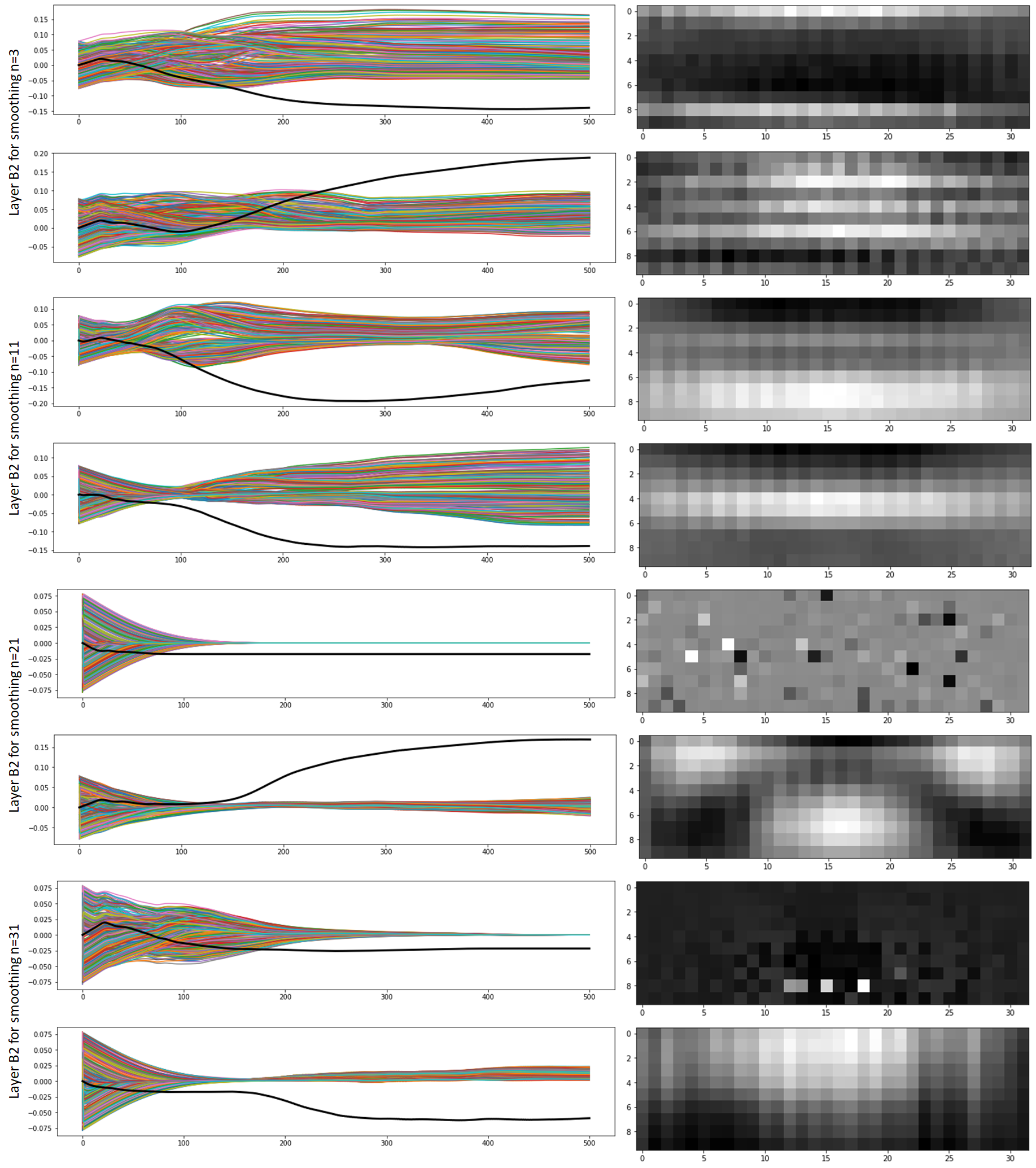}
	\caption[Resultados, pesos obtenidos en el modelo entrenado]{Weights of the B2 layers in the trained final model. On the right, the weights of the convolution kernels. On the left, the evolution of each weight during training. The thick black lines represent the biases. }
	\label{fig:res_pesos}
\end{figure*}

Figure \ref{fig:res_pesos} shows the weights of B2 and their evolution over 500 training epochs. The weights of a case where two kernels (the first for smoothing $n=21$ and the second for smoothing $n=31$) did not achieve a useful configuration are plotted intentionally. It can be seen that the aforementioned weights tend to zero because of the effect of regularization. The rest of the weights of B2 take well-defined forms, with faster magnitude variation (higher frequency) in the vertical direction (quefrency) than in the horizontal direction (time). Clearly the frequency of variation on the vertical axis decreases as $n$ increases. 

\begin{figure*}[!t]
	\centering
	\includegraphics[width=.6\textwidth]{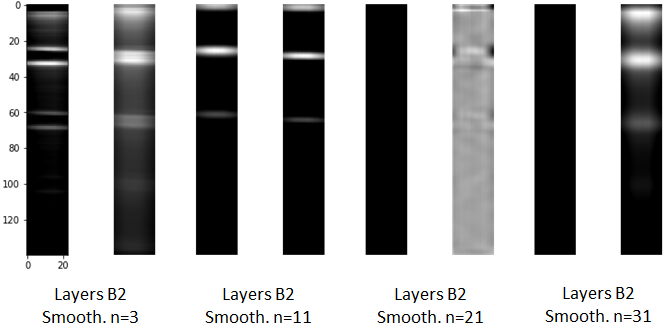}
	\caption[Resultados, salidas de capas B2 en modelo entrenado]{B2 layer outputs for figure \ref{fig:res_pesos} weights using an input with G=``0''.  }
	\label{fig:res_salidasB2}
\end{figure*}

Figure \ref{fig:res_salidasB2} shows the outputs of the B2 layers for a case where G=``0'' using the weights of figure \ref{fig:res_pesos}.

\section{Discussion}

\subsection{Metrics} \label{discu_metri}

\noindent A classification, with balanced data, carried out randomly on four classes would have an accuracy of 25\%, therefore the 71.1\% achieved shows that the neural network is capable of recognizing certain voice patterns that are related to the perception of dysphonia. 

The value of MAE and the confusion matrix (table \ref{tab:resultados_cm}) show that, for cases where there is no match, the model most of the time assigns classes close to the reference one; such as occurs among human experts, where it is unusual to differ by more than one level of G. Notice that, misclassifications (off-diagonal values in table \ref{tab:resultados_cm}) are concentrated near the diagonal. 

It is difficult to compare the results with other works, mainly because different databases are used. Some articles report accuracy greater than 80\% in the prediction of G \cite{xie2016deep, wang2016automatic,  moro2015modulation}, while the Godino-Llorente group, renowned on this topic, obtained lower accuracy and higher MAE than those of this research in two recent papers \cite{arias2019multimodal, gomez2019emulating}. A more useful analysis than direct comparison of accuracies would include the inter-rater and intra-rater agreement for each database, but unfortunately this information is not available.

In this work, for both metrics, the results obtained are very close to those that an average rater would obtain evaluating the same set of audios for the second time. Also for both metrics a great improvement can be seen comparing with the inter-rater agreement and the inter-rater error. 

Based on the results, the clinical application of the neural network obtained could be considered. On the one hand, the mean values of the metrics are close to the intra-rater values of an expert, therefore clinical use would allow professionals with less experience to obtain an automatic assessment as a reference. On the other hand, the accuracy far exceeds the agreement between human raters, therefore the widespread use of automatic classification would have a standardization effect among different professionals and health centers.

\subsection{Model architecture}

As mentioned above, few \emph{ad hoc} deep neural networks exist for this domain. Some are those of Arias-Lodoño \emph{et al.} \cite{arias2019multimodal} and Harar \emph{et al.} \cite{harar2017voice}.

In \cite{arias2019multimodal}, good results are not reported, since the deep neural network is compared with a simpler neural model that receives acoustic measurements as input, and the simpler model performs better. In \cite{harar2017voice} an \emph{ad hoc} neural network is presented to classify between healthy and pathological voices. 
Harar's model is completely different from Arias-Lodoño's. The results of \cite{harar2017voice} should be revised because the data is not well balanced. 

In this context, where there is no knowledge about the characteristics that a deep neural network must have to classify vocal quality, the developments of this work can provide useful information. 

First of all, the feature extraction layers divided into two paths, one to obtain information about noise and the other about amplitude and frequency perturbations, are novel and, according to the results, fulfill their purpose. The idea of two separate paths with convolutional layers was also used in the work of Arias-Lodoño, although with different objectives; it was used to obtain separate acoustic and modulation features on modulation spectra inputs.

Regarding the use of more than one smoothing scale, the effect may be related to a well-known concept in image pattern recognition, the scalar space theory. This proposes that the human eye is capable of recognizing patterns on different smoothing levels (at the same time) of the original image and that patterns recognized at different levels improve the overall classification. There are works on audio pattern recognition that use the same idea  \cite{von2019multi, zhu2016learning, chi2006spectrum, mesgarani2006discrimination}, but not on vocal quality.

The evolution and final configuration of the synaptic weights were analyzed, but no clear information was obtained on the internal representations or patterns in the inputs. The lack of complexity of the neural network (few kernels in this case) may be related to the difficulty in finding these patterns. No relationships were found between the presence of inputs belonging to certain classes and activation zones of neurons in the convolution layers. It was noticed that the activations of the B2 layers decrease slightly as G increases. This could indicate that some kernels are detecting the existing regularities in normal voices. 

\subsection{Limitations}

Due to the limited amount of data available, more complex models could not be analyzed. It is possible that by increasing the complexity of the neural network, such as increasing the number of kernels in the feature extraction layers, the model improves its performance. 

In the last paragraph of section \ref{discu_metri} the possibility of achieving clinical applications based on the presented neural network is mentioned. Although this model is thought to be a good starting point for future work, it should not be interpreted that the final model presented can be used directly in clinical practice. This is a frequent situation in artificial intelligence applied to medicine. Despite the efforts, most of the advances that artificial intelligence and deep learning achieve in scientific works are not applied in clinical practice. 
This topic is widely studied in the area of automatic medical image processing. The challenges for artificial intelligence to achieve clinical value have been addressed in numerous works \cite{colling2019artificial, acs2020artificial, reinke2021common, van2021deep, yoshida2021requirements, kobayashi2021state}. Many of these challenges are common to those found in  vocal pathology. 

An important and very common problem is that the data belongs to a specific population. A clear example is the work of Kather et al.\cite{kather2019deep}, where deep learning is used to detect microsatellite instability on images. An $\textnormal{AUC ROC} = 0.84$ was achieved on validation data. The same model achieved an $\textnormal{AUC ROC} < 0.69$ when was tested with a Japanese database, likely because the training data was 80\% non-Asian. This indicates the need for multicenter training data to obtain more general classification models. In addition to patient differences, samples from different laboratories have great variability in staining, image quality, scanning, and tissue preparation.

Regarding vocal quality, to achieve a general classifier, it will be necessary to have data from different ages, occupations, gender, languages and geographic regions, as well as variability in recording conditions (for example different environments, microphones and sound cards). 

Obtaining recordings with the necessary variability is a difficult task, but so is the classification of all voices by a sufficient number of specialists (at least twice each to measure inter-rater agreement). The main challenge in achieving the application of artificial intelligence in the practice of vocal clinic is probably related to achieving more general and larger databases to fit the models. Then, the use of this data will enable or force changes in the neural network architecture.

\section{Conclusion}
\noindent 

It is concluded that the obtained model is capable of predicting G, on PVQD, with an accuracy close to the accuracy of a human evaluator.

Regarding the architecture, particularly on feature extraction, it is concluded that a deep neural network designed to recognize amplitude perturbation, frequency perturbation and noise patterns obtains useful information for G prediction.

Finally, it is concluded that the presented model can be a starting point for future development of voice quality classifiers, which should be trained and evaluated with larger databases containing greater variability to achieve their application in clinical practice.

%Bibliography
\bibliographystyle{unsrt}  
\bibliography{article}

\begin{thebibliography}{10}

\bibitem{barsties2015assessment}
Ben Barsties and Marc De~Bodt.
\newblock Assessment of voice quality: current state-of-the-art.
\newblock {\em Auris Nasus Larynx}, 42(3):183--188, 2015.

\bibitem{hazan2012early}
Hananel Hazan, Dan Hilu, Larry Manevitz, Lorraine~O Ramig, and Shimon Sapir.
\newblock Early diagnosis of parkinson's disease via machine learning on speech
  data.
\newblock In {\em 2012 IEEE 27th Convention of Electrical and Electronics
  Engineers in Israel}, pages 1--4. IEEE, 2012.

\bibitem{little2008suitability}
Max Little, Patrick McSharry, Eric Hunter, Jennifer Spielman, and Lorraine
  Ramig.
\newblock Suitability of dysphonia measurements for telemonitoring of
  parkinson’s disease.
\newblock {\em Nature Precedings}, pages 1--1, 2008.

\bibitem{isshiki1966approach}
N~Isshiki, N~Yanagihara, and M~Morimoto.
\newblock Approach to the objective diagnosis of hoarseness.
\newblock {\em Folia Phoniatrica et Logopaedica}, 18(6):393--400, 1966.

\bibitem{hirano1986clinical}
Minoru Hirano and Karen~R McCormick.
\newblock Clinical examination of voice by minoru hirano, 1986.

\bibitem{yun2005correlation}
Young~Sun Yun, Eun~Kyung Lee, Chung~Hwan Baek, and Young~Ik Son.
\newblock The correlation of grbas scales and laryngeal stroboscopic findings
  for the assessment of voice therapy outcome in the patients with vocal
  nodules.
\newblock {\em Korean Journal of Otolaryngology-Head and Neck Surgery},
  48(12):1501--1505, 2005.

\bibitem{hui2007validation}
Huangfu Hui, Kong Weijia, Gong Shusheng, et~al.
\newblock The validation of acoustic analysis and subjective judgment scales of
  several voice disorders [j].
\newblock {\em Journal of Audiology and Speech Pathology}, 3(010), 2007.

\bibitem{karnell2007reliability}
Michael~P Karnell, Sarah~D Melton, Jana~M Childes, Todd~C Coleman, Scott~A
  Dailey, and Henry~T Hoffman.
\newblock Reliability of clinician-based (grbas and cape-v) and patient-based
  (v-rqol and ipvi) documentation of voice disorders.
\newblock {\em Journal of Voice}, 21(5):576--590, 2007.

\bibitem{jesus2009voice}
Luis~MT Jesus, Anna Barney, Pedro S{\'a}~Couto, Helena Vilarinho, and Ana
  Correia.
\newblock Voice quality evaluation using cape-v and grbas in european
  portuguese.
\newblock In {\em Sixth International Workshop on Models and Analysis of Vocal
  Emissions for Biomedical Applications}, 2009.

\bibitem{kreiman2010perceptual}
Jody Kreiman and Bruce~R Gerratt.
\newblock Perceptual assessment of voice quality: past, present, and future.
\newblock {\em Perspectives on Voice and Voice Disorders}, 20(2):62--67, 2010.

\bibitem{nunez2012espectrograma}
Faustino N{\'u}nez-Batalla, Juan~Pablo D{\'\i}az-Molina, Isabel
  Garc{\'\i}a-L{\'o}pez, Adriana Moreno-M{\'e}ndez, Mar{\'\i}a Costales-Marcos,
  Carla Moreno-Galindo, and Pablo Mart{\'\i}nez-Camblor.
\newblock El espectrograma de banda estrecha como ayuda para el aprendizaje del
  m{\'e}todo grabs de an{\'a}lisis perceptual de la disfon{\'\i}a.
\newblock {\em Acta Otorrinolaringol{\'o}gica Espa{\~n}ola}, 63(3):173--179,
  2012.

\bibitem{gordillo2018hitos}
Luisa Fernanda~Angel Gordillo.
\newblock Hitos de la evaluaci{\'o}n perceptual auditiva de la voz: ?`hay
  evidencia?
\newblock {\em Aret{\'e}}, 18(2):65--74, 2018.

\bibitem{kreiman1998validity}
Jody Kreiman and Bruce~R Gerratt.
\newblock Validity of rating scale measures of voice quality.
\newblock {\em The Journal of the Acoustical Society of America},
  104(3):1598--1608, 1998.

\bibitem{huang2017densely}
Gao Huang, Zhuang Liu, Laurens Van Der~Maaten, and Kilian~Q Weinberger.
\newblock Densely connected convolutional networks.
\newblock In {\em Proceedings of the IEEE conference on computer vision and
  pattern recognition}, pages 4700--4708, 2017.

\bibitem{muhammad2018edge}
Ghulam Muhammad, Mohammed~F Alhamid, Mansour Alsulaiman, and Brij Gupta.
\newblock Edge computing with cloud for voice disorder assessment and
  treatment.
\newblock {\em IEEE Communications Magazine}, 56(4):60--65, 2018.

\bibitem{alhussein2018voice}
Musaed Alhussein and Ghulam Muhammad.
\newblock Voice pathology detection using deep learning on mobile healthcare
  framework.
\newblock {\em IEEE Access}, 6:41034--41041, 2018.

\bibitem{xie2016deep}
Simin Xie, Nan Yan, Ping Yu, Manwa~L Ng, Lan Wang, and Zhuanzhuan Ji.
\newblock Deep neural networks for voice quality assessment based on the grbas
  scale.
\newblock {\em Interspeech 2016}, pages 2656--2660, 2016.

\bibitem{fang2019detection}
Shih-Hau Fang, Yu~Tsao, Min-Jing Hsiao, Ji-Ying Chen, Ying-Hui Lai, Feng-Chuan
  Lin, and Chi-Te Wang.
\newblock Detection of pathological voice using cepstrum vectors: A deep
  learning approach.
\newblock {\em Journal of Voice}, 33(5):634--641, 2019.

\bibitem{arias2019multimodal}
Juli{\'a}n~D Arias-Londo{\~n}o, Jorge~A G{\'o}mez-Garc{\'\i}a, and Juan~I
  Godino-Llorente.
\newblock Multimodal and multi-output deep learning architectures for the
  automatic assessment of voice quality using the grb scale.
\newblock {\em IEEE Journal of Selected Topics in Signal Processing},
  14(2):413--422, 2019.

\bibitem{harar2017voice}
Pavol Harar, Jesus~B Alonso-Hernandezy, Jiri Mekyska, Zoltan Galaz, Radim
  Burget, and Zdenek Smekal.
\newblock Voice pathology detection using deep learning: a preliminary study.
\newblock In {\em 2017 international conference and workshop on bioinspired
  intelligence (IWOBI)}, pages 1--4. IEEE, 2017.

\bibitem{kreiman1996perceptual}
Jody Kreiman and Bruce~R Gerratt.
\newblock The perceptual structure of pathologic voice quality.
\newblock {\em The Journal of the Acoustical Society of America},
  100(3):1787--1795, 1996.

\bibitem{walden2020perceptual}
Patrick Walden.
\newblock Perceptual voice qualities database (pvqd): Database characteristics.
\newblock {\em Journal of Voice: Official Journal of the Voice Foundation},
  2020.

\bibitem{godino2003effects}
Juan~Ignacio Godino-Llorente, Tim Ritchings, and Carl Berry.
\newblock The effects of inter and intra speaker variability on pathological
  voice quality assessment.
\newblock In {\em Third International Workshop on Models and Analysis of Vocal
  Emissions for Biomedical Applications}, 2003.

\bibitem{ritchings2002pathological}
RT~Ritchings, Mark McGillion, and CJ~Moore.
\newblock Pathological voice quality assessment using artificial neural
  networks.
\newblock {\em Medical engineering \& physics}, 24(7-8):561--564, 2002.

\bibitem{garcia2021deep}
Mario~Alejandro García.
\newblock {\em Clasificación automática del grado general de disfonía}.
\newblock PhD thesis, Universidad Tecnológica Nacional, 12 2021.

\bibitem{wang2016automatic}
Zhijian Wang, Ping Yu, Nan Yan, Lan Wang, and Manwa~L Ng.
\newblock Automatic assessment of pathological voice quality using
  multidimensional acoustic analysis based on the grbas scale.
\newblock {\em Journal of Signal Processing Systems}, 82(2):241--251, 2016.

\bibitem{moro2015modulation}
Laureano Moro-Vel{\'a}zquez, Jorge~Andr{\'e}s G{\'o}mez-Garc{\'\i}a,
  Juan~Ignacio Godino-Llorente, and Gustavo Andrade-Miranda.
\newblock Modulation spectra morphological parameters: a new method to assess
  voice pathologies according to the grbas scale.
\newblock {\em BioMed research international}, 2015, 2015.

\bibitem{gomez2019emulating}
Jorge G{\'o}mez-Garc{\'\i}a, Laureano Moro-Vel{\'a}zquez, Janaina
  Mendes-Laureano, Germ{\'a}n Castellanos-Dominguez, and Juan~Ignacio
  Godino-Llorente.
\newblock Emulating the perceptual capabilities of a human evaluator to map the
  grb scale for the assessment of voice disorders.
\newblock {\em Engineering Applications of Artificial Intelligence},
  82:236--251, 2019.

\bibitem{von2019multi}
Patrick von Platen, Chao Zhang, and Philip Woodland.
\newblock Multi-span acoustic modelling using raw waveform signals.
\newblock {\em Proc. Interspeech 2019}, pages 1393--1397, 2019.

\bibitem{zhu2016learning}
Zhenyao Zhu, Jesse~H Engel, and Awni Hannun.
\newblock Learning multiscale features directly from waveforms.
\newblock {\em Interspeech 2016}, pages 1305--1309, 2016.

\bibitem{chi2006spectrum}
Taishih Chi and Shihab~A Shamma.
\newblock Spectrum restoration from multiscale auditory phase singularities by
  generalized projections.
\newblock {\em IEEE transactions on audio, speech, and language processing},
  14(4):1179--1192, 2006.

\bibitem{mesgarani2006discrimination}
Nima Mesgarani, Malcolm Slaney, and Shihab~A Shamma.
\newblock Discrimination of speech from nonspeech based on multiscale
  spectro-temporal modulations.
\newblock {\em IEEE Transactions on Audio, Speech, and Language Processing},
  14(3):920--930, 2006.

\bibitem{colling2019artificial}
Richard Colling, Helen Pitman, Karin Oien, Nasir Rajpoot, Philip Macklin,
  CM-Path~AI in~Histopathology Working~Group, Velicia Bachtiar, Richard Booth,
  Alyson Bryant, Joshua Bull, et~al.
\newblock Artificial intelligence in digital pathology: a roadmap to routine
  use in clinical practice.
\newblock {\em The Journal of pathology}, 249(2):143--150, 2019.

\bibitem{acs2020artificial}
Bal{\'a}zs Acs, Mattias Rantalainen, and Johan Hartman.
\newblock Artificial intelligence as the next step towards precision pathology.
\newblock {\em Journal of internal medicine}, 288(1):62--81, 2020.

\bibitem{reinke2021common}
Annika Reinke, Matthias Eisenmann, Minu~D Tizabi, Carole~H Sudre, Tim
  R{\"a}dsch, Michela Antonelli, Tal Arbel, Spyridon Bakas, M~Jorge Cardoso,
  Veronika Cheplygina, et~al.
\newblock Common limitations of image processing metrics: A picture story.
\newblock {\em arXiv preprint arXiv:2104.05642}, 2021.

\bibitem{van2021deep}
Jeroen van~der Laak, Geert Litjens, and Francesco Ciompi.
\newblock Deep learning in histopathology: the path to the clinic.
\newblock {\em Nature medicine}, 27(5):775--784, 2021.

\bibitem{yoshida2021requirements}
Hiroshi Yoshida and Tomoharu Kiyuna.
\newblock Requirements for implementation of artificial intelligence in the
  practice of gastrointestinal pathology.
\newblock {\em World Journal of Gastroenterology}, 27(21):2818, 2021.

\bibitem{kobayashi2021state}
Soma Kobayashi, Joel~H Saltz, and Vincent~W Yang.
\newblock State of machine and deep learning in histopathological applications
  in digestive diseases.
\newblock {\em World Journal of Gastroenterology}, 27(20):2545, 2021.

\bibitem{kather2019deep}
Jakob~Nikolas Kather, Alexander~T Pearson, Niels Halama, Dirk J{\"a}ger,
  Jeremias Krause, Sven~H Loosen, Alexander Marx, Peter Boor, Frank Tacke,
  Ulf~Peter Neumann, et~al.
\newblock Deep learning can predict microsatellite instability directly from
  histology in gastrointestinal cancer.
\newblock {\em Nature medicine}, 25(7):1054--1056, 2019.

\end{thebibliography}

\end{document}